\documentclass{aastex}
\usepackage{natbib,emulateapj5,apjfonts}

\def\omegatot{\Omega_{\rm tot}}

\begin{document}

\journalinfo{Submitted to The Astrophysical Journal}
\title{First Intrinsic Anisotropy Observations with the Cosmic Background 
Imager}

\author{S. Padin, J. K. Cartwright, B. S. Mason, T. J. Pearson, A. C. S. 
Readhead,\\
M. C.  Shepherd, J.~Sievers, and P. S. Udomprasert}

\affil{California Institute of Technology, 1200 East California 
Boulevard, Pasadena, CA 91125}

\author{W. L. Holzapfel}

\affil{University of California, 426 LeConte Hall, Berkeley, CA 94720-7300}

\author{S. T. Myers}

\affil{National Radio Astronomy Observatory, P.O. Box O, Socorro, NM 87801}

\author{J. E. Carlstrom and E. M. Leitch}

\affil{University of Chicago, 5640 South Ellis Ave., Chicago, IL 60637}

\author{M. Joy}

\affil{Dept. of Space Science, SD50, NASA Marshall Space Flight Center, 
Huntsville, AL 
35812}

\and
\author{L. Bronfman and J. May}

\affil{Departamento de Astronom\'{\i}a, Universidad de Chile, Casilla 36-D, 
Santiago, 
Chile}

\begin{abstract}
We present the first results of observations of the intrinsic
anisotropy of the cosmic microwave background radiation with the
Cosmic Background Imager from a site at 5080 m altitude in northern
Chile.  Our observations show a sharp decrease in $C_l$ in the range
$l=400$--$1500$. The broadband amplitudes we have measured are $\delta
T_{\rm band}=58.7_{-6.3}^{+7.7}\; \mu$K for $l=603_{-166}^{+180}$ and
$\delta T_{\rm band}=29.7_{-4.2}^{+4.8}\; \mu$K for
$l=1190_{-224}^{+261}$, where these are half-power widths in $l$.
Such a decrease in power at high $l$ is one of the fundamental
predictions of the standard cosmological model, and these are the
first observations which cover a broad enough $l$ range to show this
decrease in a single experiment.  The $C_l$ we have measured enables us
to place limits on the density parameter, $\omegatot\le 0.4$ or
$\omegatot \ge 0.7$ (90\% confidence).
\end{abstract}


\section{Introduction}

In standard cosmologies, spatial temperature variations in the cosmic
microwave background radiation (CMBR) are closely related to the
primordial density fluctuations which gave rise to the formation of
all structure in the universe \citep{PEEBLES,SUNYAEV}.  The angular
power spectrum of these temperature variations on the celestial
sphere, $C_l$, yields a direct estimate of the prime cosmological
parameters and provides a fundamental link between particle physics
and cosmology \citep[e.g.,][]{KAMION}.  Since radio interferometers
sample the angular power spectrum directly and are straightforward to
calibrate, they provide a simple and direct determination of
$C_l$. Here we report the first observations of the CMBR with the
Cosmic Background Imager (CBI).

\section{The Cosmic Background Imager}

The CBI is a radio interferometer with 13 0.9~m diameter
antennas mounted on a 6~m tracking platform.  It operates in 10
1~GHz frequency channels from 26 to 36 GHz.  The instantaneous
field of view and the maximum resolution are $\sim 45'$ and $\sim 3'$
(FWHM).  The instrument has an altitude-azimuth mount, and the antenna
platform can also be rotated about the optical axis to increase the
aperture-plane $(u,v)$ coverage and to facilitate polarization
observations.  The antennas have low-noise broadband high-electron mobility transistor (HEMT) amplifier
receivers with $\sim25$ K noise temperatures.  The typical system
noise temperature averaged over all 10 bands is $\sim30$ K, including
ground spillover and atmosphere.  The frequency of operation of the
CBI was chosen as a compromise between the effects of astronomical
foregrounds, atmospheric emission, and the sensitivity that can be
achieved with HEMT amplifiers.  Details of the instrument design may
be found in \citet[][S. Padin et al. 2000, in preparation]{Padin1, Padin2} and on the CBI web
 site.\footnote{See 
\url{http://www.astro.caltech.edu/$\sim$tjp/CBI.}} The CBI is located
at an altitude of 5080 m near Cerro Chajnantor in the Atacama desert
in northern Chile.  This site was chosen because the atmospheric
opacity is low and the CBI can operate at the thermal noise limit much
of the time. The instrument was assembled and tested on the Caltech
campus during 1998 and 1999 and shipped to Chile in 1999 August.
Installation of the telescope and site infrastructure were completed
by the end of 1999, and the full instrument has been in operation
since early 2000 January.

The CBI is sensitive to multipoles in the range $400<l<4250$, where
these values reflect the half-power widths of the window functions on
the shortest and longest baselines.  The CBI complements BOOMERANG \citep{BOOM1, BOOM2}, the Degree Angular Scale Inteferometer (DASI) \citep{DASI},
{\it Microwave Anisotropy Probe},\footnote{See the {\it MAP} Web site at
\url{http://map.gsfc.nasa.gov/}.} MAXIMA \citep{Maxima1, Maxima2},
and the Very Small Array \citep{VSA}, which cover the range $50<l<1000$.  DASI 
is a
sister project to the CBI, and the CBI and DASI designs were chosen to
complement each other.  The CBI control software and
correlator and receiver control electronics were duplicated by the DASI team for the DASI project.  Together these two interferometers cover
the multipole range $100<l<4250$.

\section{Observations}

The antenna platform of the CBI permits a wide variety of antenna
configurations.  For observations during the test phase
(2000 January--April) we chose a configuration with the antennas
around the perimeter of the platform, which provided easy access to
the receivers and fairly uniform $(u,v)$ coverage, enabling us to
test the full range of CBI baselines.  We report here only
observations on baselines corresponding to $l<1510$, which account for
25\% of the data in this ring configuration. We do not report on the
higher multipole bins because these are more dependent on the 
bright-source and statistical faint-source corrections, which are still
preliminary.  These results, and mosaicked observations which
significantly increase the resolution in $l$, will be presented
elsewhere.

We based our flux density scale on observations of Jupiter, assuming $T_{\rm Jupiter} = 152$ K at 32 GHz, with 5\% uncertainty
\citep{MASONCAL}.  The spectral index of Jupiter is not constant between
26 and 36 GHz \citep{Wrixon71}, so we used Taurus A as our prime calibrator.  Taurus A is slightly resolved with the CBI,
but it can be well fitted by an elliptical Gaussian model.  We referenced the 32 GHz
flux density of Taurus A to that of Jupiter and transferred this
to the other frequency channels assuming $\alpha_{\rm Taurus~A}=-0.30$, where $S\propto \nu^{\alpha}$ \citep{Mezger86}. Any uncertainties resulting from this extrapolation are $< 1\%$.

For our intrinsic anisotropy observations, we selected regions at
Galactic latitudes above $20^{\circ}$, for which the synchrotron
emission and the IRAS $100\; \mu$m emission are relatively low, and
which avoided bright point sources detected in the NRAO Very Large Array 1.4 GHz
Sky Survey (NVSS) \citep{NVSS}.  These regions are also in the declination
range $-5^{\circ} < \delta < -2^{\circ}$, which permits point-source
monitoring with the Owens Valley Radio Observatory (OVRO) 40~m
telescope (see \S~5).

Daytime observations of the CMBR are not possible because radio
emission from the Sun contaminates the visibilities.  Radio emission
from the Moon causes similar problems, and intrinsic anisotropy
observations are restricted to fields more than $60^{\circ}$ from the
moon. The weather during 2000 January--April was uncommonly poor, and
we lost 50\% of the nights as a result of bad weather.  In the remaining
nights the observing conditions were superb, and the sensitivity was
limited by the system noise.

The dominant systematic contamination in the CBI observations is due
to ground spillover.  The level of contamination on the 1~m
baselines is typically between a few tens and a few hundreds of mJy;
in rare instances it can be as high as a few Jy, and it falls off with
increasing baseline length.  The ground signals can be recognized in
the maps because they are not confined to the primary beam area, as
celestial signals are.  We have successfully removed the ground signal
by differencing visibilities measured on two fields at the same
declination separated by $08^{\rm m}$ in right ascension.  The observations
were alternated between two fields on an 8~minute timescale, so that both
fields were observed at the same position relative to the ground.  In
addition to filtering out ground spillover, this differencing strongly
rejects crosstalk and other spurious instrumental signals. As
discussed below, any residual spurious signals amount to
$\lesssim\,1.3$\% of $C_l^{1/2}$ and may therefore safely be ignored. The
differenced observations reported here are between two fields centered
at $08^{\rm h}\,44^{\rm m}\,40^{\rm s}$, $ -03^{\circ}\,10'$ and
$08^{\rm h}\,52^{\rm m}\,40^{\rm s}$, $ -03^{\circ}\,10'$; and
two fields centered at $14^{\rm h}\,42^{\rm m}$, $
-03^{\circ}\,50'$ and $14^{\rm h}\,50^{\rm m}$, $
-03^{\circ}\,50'$ (J2000). To search for residual spurious
signals in excess of the thermal noise level in our differenced
visibilities, we needed an accurate estimate of the thermal noise
level.  We obtained this estimate by computing the rms of successive
8.4~s integrations in the differenced visibilities.  The typical noise
level computed in this way was 2 Jy, in agreement with the (less
accurate) noise level determined from the system temperature.  The
means of the $8^{\rm m}$ scans were much smaller than 2 Jy, so this
method provided an accurate estimate of the thermal noise level.  To
test for residual spurious signals, we divided our differenced
visibilities into two parts in three ways: (1) pre- versus posttransit observations; (2) middle half versus outer quarters of hour-angle 
range; and (3) first versus second half of epochs on each field (pre--March 23 vs.\ post--March 23).

In each case, we subtracted the singly differenced visibilities
point by point in the $(u,v)$ plane.  In cases 1 and 2 the doubly
differenced visibilities were consistent with the expected noise.  In
case 3 there was an excess amounting to a 1.3\% contamination in
$C_l^{1/2}$.

\section{Analysis and Results}

We observed for 58.5~hr on each of the $08{\rm ^h}$ fields and for 16.15~hr
on each of the $14^{\rm h}$ fields.  
The sky signal is clearly visible in the differenced images on the
100 and 104~cm baselines (Fig.~1). 

\begin{figure*}
\epsfxsize=\hsize
\epsfbox{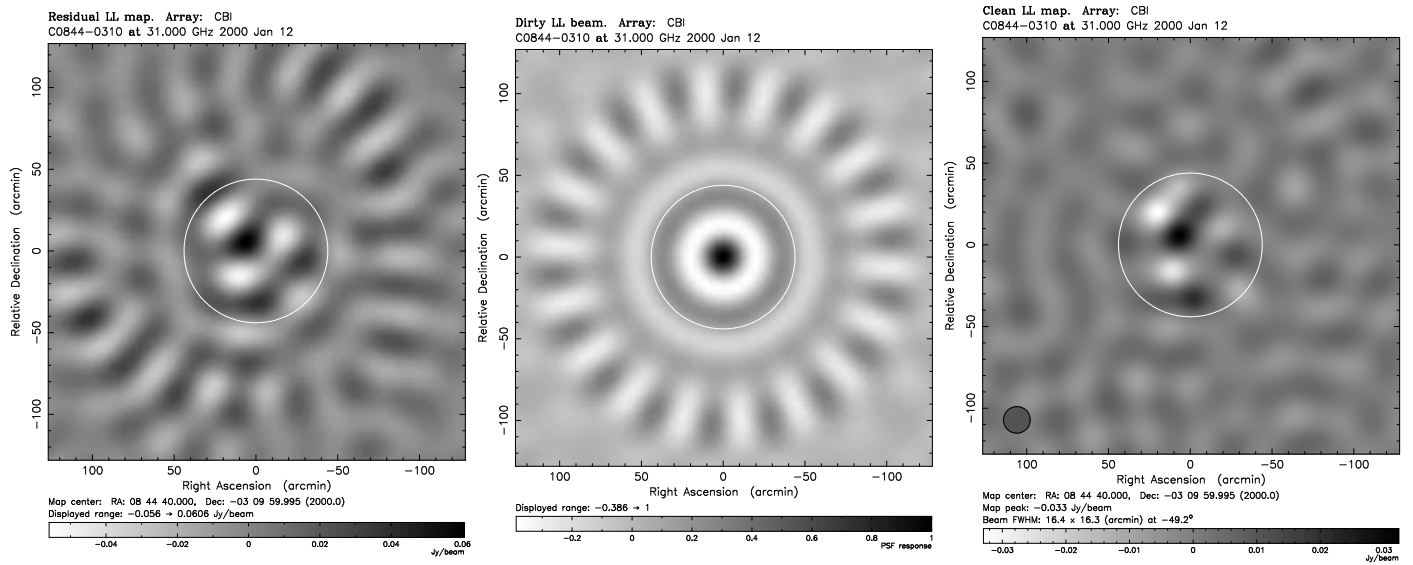}
\caption{Differenced image of the $08^{\rm h}$ field observed on the
nine 100~cm and 104~cm CBI baselines in the test configuration ({\it left}),
the corresponding point-spread function ({\it center}), and an image
deconvolved with the H\"ogbom CLEAN algorithm ({\it right}), assuming that
the emission is confined to the primary-beam area.  The 10 frequency
channels have been combined, and the synthesized beamwidth $\sim 16'$
(FWHM).  The circle at radius $44'$ indicates the primary-beam
area which was cleaned. The structures outside the circle in the
uncleaned image ({\it left}) are due to sidelobes.  These structures
disappear after cleaning within the primary-beam area, indicating that
they are caused by sidelobes.}
\end{figure*}

The extraction of the angular spectrum from visibility measurements is
straightforward \citep{White99}.  The covariance matrix of the
observations is the matrix of the covariances between all the
visibility measurements:
\begin{equation}
C\;=\;M\,+\,N,
\end{equation}
where $M$ and $N$ are the sky and noise covariance
matrices.  We assume that the noise on different baselines and at
different frequencies is uncorrelated, i.e., that $N$ is diagonal.  The
sky covariance matrix is 
\begin{eqnarray}
M_{jk}&=&\langle V({\bf u}_j,\nu_j)V^*({\bf
u}_k, \nu_k) \rangle\\
&=&\int \int d^2{\bf v} \tilde A({\bf u}_j -
{\bf v}, \nu_j)\tilde A^*({\bf u}_k - {\bf v}, \nu_k)S({\bf v},
\nu_j,\nu_k)
\end{eqnarray}
for two visibility points $j,k$, where $\tilde
A({\bf u},\nu)$ is the Fourier transform of the primary beam at
frequency $\nu$, ${\bf u}$ is the baseline vector in wavelengths, and
$S({\bf v}, \nu_j, \nu_k)$ is a generalized power spectrum of the
intensity fluctuations \citep{Hobson95}. The effective weighting
of this power spectrum defines the window function.  The indices, $j$
and $k$, run from 1 to $n$ where $n$ is the number of distinct $(u,v)$
points.  We average all the data taken at different times for each
$(u,v)$ point before doing the maximum likelihood calculation.

The generalized power spectrum is related to $C_l$ by
\begin{equation}
S( v, \nu_j,\nu_k) = \left(2kT_0 \over c^2 \right)^2\, 
\nu_j^2\,\nu_k^2\,g(\nu_j)\,g(\nu_k)\,C_l ,
\end{equation}
where $l+{1 \over 2}\,=\,2\pi |{\bf v}|$ and the $g$ factor is a small 
correction for the difference 
between the Rayleigh-Jeans and Planck
functions. 
We can test a hypothetical power spectrum,
$[C_l]$, by
forming the likelihood function
\begin{equation}
L([C_l])= {1 \over {\pi^n \, {\rm det}\,C}}\; {\rm exp}\,\{-V^*({\bf
u}_j)C_{jk}^{-1}V({\bf u}_k)\} .
\end{equation}
The cross-correlation between the signals received from two fields
separated by $08^{\rm m}$ in R.A. is negligible, so the expected variance of the
differenced visibilities is twice the variance of the undifferenced
visibilities. Our parametric model for $C_l$ consists of two
parameters, these being the amplitudes of $C_l$ in the two ranges
$l<900$ and $l>900$, assuming $l(l+1)C_l$ is constant in each
range. In the test configuration there is a gap in our $(u,v)$
coverage between $l=800$ and $l=1000$ which makes this a natural
division.  The band-power window functions, here approximated by sums
of the single-baseline single-channel window functions within each
$l$ bin, may be characterized by $l=603_{-166}^{+180}$ and
$l=1190_{-224}^{+261}$ (half-power widths).  The maximum likelihood
broadband signals we measure are $\delta T_{\rm
band}\equiv[l(l+1)C_l/(2\pi)]^{1/2}\times T_{\rm
cmb}=58.7_{-6.3}^{+7.7}\;\mu$K for the $l=603$ bin and $\delta T_{\rm
band}=29.7_{-4.2}^{+4.8}\;\mu$K for the $l=1190$ bin.  The error bars
indicate the points at which the likelihood has dropped by $e^{-0.5}$
(which are within 10\% of the 68\% integrated probability values).
The results for the two fields, shown in Figure 2, agree to within the
uncertainties. For the lower $l$ bin, the uncertainties are dominated
by sample variance, and they would be decreased by $<2\%$ in the
absence of thermal noise.  For the upper $l$ bin, the uncertainties
would be decreased by 31\% and 54\% for the $08^{\rm h}$ and $14^{\rm
h}$ fields, respectively, in the absence of thermal noise.

\begin{figure*}
\plotone{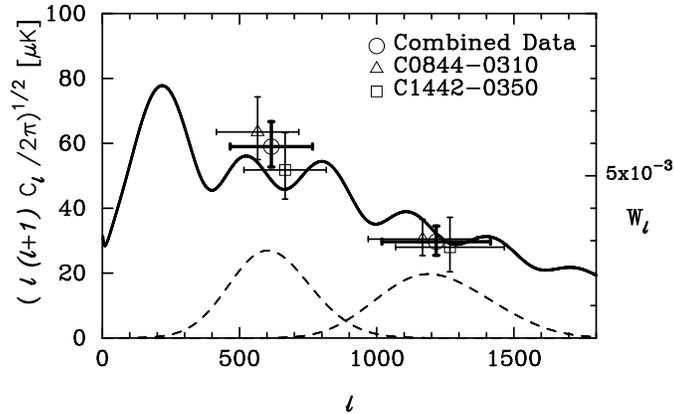}
\caption{CMBR anisotropy spectrum determined from CBI
observations.  The triangles and squares show results on the $08^{\rm
h}$ and $14^{\rm h}$ differenced fields; the circles show the results
of a joint maximum likelihood analysis of both differenced fields.
The individual $08^{\rm h}$ and $14^{\rm h}$ field results are offset
in $l$ for clarity.  The window functions for each bin are shown as
dashed lines. The solid curve represents a flat model universe with
$H_0=75\;{\rm km\;s^{-1}\;Mpc^{-1}},\;\Omega_{\rm b}h^2=0.019$,
 and $\Omega_{\rm cdm}=0.2$.}
\end{figure*}

Our maximum likelihood analysis has been tested using software written
independently by two of the authors, with no common code between the
packages and using significantly different implementations for
important steps, such as evaluation of the window function, binning
and maximization algorithm.  We have analyzed both the real data and
simulated differenced data sets generated by realizations of known
power spectra together with realistic noise and point sources. The
results from these two software packages are in excellent agreement,
and the simulations recover the original input power spectra. We have
also generated 54 simulations of differenced sky images based on  our
observed band powers in the two bins (Fig.~2) and compared the rms
signal, measured within the primary-beam area in these simulations,
with the rms fluctuations in the primary beam measured in actual
observations of 54 differenced fields, to be published elsewhere.
Both the means and the distributions of the rms values for the
observed and simulated fields are in excellent agreement.  Thus, we
are confident that our derived spectrum is a reliable representation
of the signal that we have detected on the sky.

\section{Foregrounds}

Radio galaxies and radio-loud quasars are a source of confusion at CBI
frequencies and angular scales, so we equipped the OVRO 40~m telescope
with a four-channel 26--34 GHz receiver for point-source monitoring,
and observed all of the sources in the NVSS with $S_{1.4 \,{\rm
GHz}}>6$ mJy in our CBI fields.  Those sources detected with the 40~m
telescope at the $3\,\sigma$ level ($S_{30\, {\rm GHz}}\geq 6$ mJy)
have been subtracted from our CBI visibility data using the flux
densities measured on the 40~m telescope.  This reduced the levels of
$\delta T_{\rm band}$ measured in the lower and upper $l$ bins by
0.5\% and 1\%.  We have also applied corrections based on the source
count statistics \citep{FIRST} to account for point sources with
$S_{30\, {\rm GHz}}<6$ mJy, which have not been subtracted
individually from our visibility data.  The corrections amounted to
decreases in $\delta T_{\rm band}$ of 1.6\% and 8.3\% in the lower and
upper $l$ bins. These corrections have been applied to the band powers
given in \S~4.  The uncertainty in the statistical correction, $\sim
20\%$, makes $\ll0.1$ K difference to the errors in both bins.

It is unlikely that diffuse Galactic foreground emission is a
significant contaminant in our observations.  The expected rms
fluctuations due to Galactic synchrotron emission on angular scales
$5'$--$30'$ are $<9 \; \mu$K \citep[e.g.,][]{TEGMARK1}. In the RING5M
experiment, \citet{RING5M1} detected Galactic emission at 14.5 and 32
GHz with a spectrum consistent with free-free radiation but a much
higher level than predicted from H$\alpha$ measurements.  We have
therefore made 14.5 GHz observations with the OVRO 40~m telescope
along a strip at declination $-5^{\circ}$ over the right ascension
range $0^{\rm h}$--$24^{\rm h}$.  The beam and beamthrow, 7\farcm4 and
22\farcm2, are fairly well matched to the CBI angular scales in the
lower $l$ bin, so, after correcting for the window function, we may
use these observations to estimate the possible level of contamination
in our CBI observations.  If all of the signal seen at 14.5 GHz at
$|b^{\rm II}|>5^{\circ}$ is attributed to anomalous emission with the
same spectral properties as seen in the RING5m data, then this amounts
to a component in our CBI observations $\delta T_{\rm band} = 20$
$\mu$K due to anomalous foregrounds.  Subtraction in quadrature from
the signal we have detected in the first $l$ bin would reduce our
observed $\delta T_{\rm band}$ by 7\%.

We have measured the temperature spectral index,
$\beta=\ln(T_1/T_2)/\ln(\nu_1/\nu_2)$, with 1 month of data from the
CBI in a more compact configuration, optimized to measure both the
angular spectrum and the radio-frequency spectrum of the CMBR.  We
find that the signal is primarily CMBR and not Galactic.  We used a
maximum likelihood analysis with $\beta$ as a free parameter to
determine that $\beta = 0.0 \pm 0.4$ ($1\,\sigma$ error) in the lower
$l$ bin. If as much as 21\% of the $\delta T_{\rm band}$ in this $l$
bin were due to a free-free foreground component with spectral index
$\beta=-2.1$, while the remainder was CMBR with $\beta=0$, the
spectral index measured would be $< -0.8$, which is ruled out at the
$2\,\sigma$ level. A 15\% synchrotron foreground component with
spectral index $\beta=-2.7$ can be ruled out at the same level.

\section{Discussion}

A decrease in $C_l$ at high $l$, caused by photon diffusion and the
thickness of the last scattering region, is a fundamental prediction
of the standard cosmological model \citep{Silk}, and this is the first
time that such a decrease has been detected in a single experiment. It
is also the first time that anisotropy has been detected at
$l>1000$. The levels of $\delta T_{\rm band}$ detected with the CBI
are consistent with observations at high $l$ made over the last 12
years \citep{NCP, CAT1, SUZIE, CAT2, RING5M2, BIMA, ATCA}. The level
of $\delta T_{\rm band}$ we measure at $l \sim 600$ is a factor 1.5
higher than that found by BOOMERANG and a factor 1.3 higher than that
found by MAXIMA; here we have used the best fit spectrum to the
BOOMERANG+MAXIMA+DMR data \citep{Jaffe} to extrapolate the BOOMERANG
and MAXIMA data.  The additional power $\delta T_{\rm band}^2$
detected by the CBI is significant at the $1.8 \,\sigma$ level
(BOOMERANG) and the $1.4 \,\sigma$ level (MAXIMA), where $\sigma$
includes calibration uncertainties of 10\% (BOOMERANG), 4\% (MAXIMA)
and 5\% (CBI), pointing uncertainties of 11\% (BOOMERANG) and 5\%
(MAXIMA), the uncertainty of 13\% in the $\delta T_{\rm band}$ measured
in the lower $l$ bin (CBI),  and an estimated uncertainty of 8\% in the
$\delta T_{\rm band}$ measured by BOOMERANG and MAXIMA in the
multipole range $300<l<700$.  It is important to determine whether
these differences between the CBI and BOOMERANG-MAXIMA are real. The
RING5M experiment \citep{RING5M2} reported $\delta T_{\rm
band}=59_{-6.5}^{+8.6}\, \mu$K at $l\sim$ 600, which agrees well with
the CBI value and is discrepant at the $1.8\,\sigma$ level with
BOOMERANG and at the $1.4\,\sigma$ level with MAXIMA.  The CAT values
\citep{CAT1, CAT2} are intermediate between the CBI and BOOMERANG and
MAXIMA values, but any differences are significant only at the $\sim
1\,\sigma$ level.

\begin{table*}
\begin{center}
\caption{Parameter Space for Cosmological Likelihood Analysis}
\begin{tabular}{ l l l }
\tableline\tableline
                   & Flat Models                      & Open Models \\ 
\tableline
$\Omega_{\rm b}\, h^2$     & $0.003 \rightarrow 0.03 \, (0.003)$ & $0.01 
\rightarrow 
0.03 \, (0.01)$ \\ 
$\Omega_{\rm cdm}$     & $0.1 \rightarrow 0.5 \, (0.04)$     &
                   $\Omega_{\rm m} - \Omega_{\rm b}$ \\
$\Omega_{\rm m}$         & $\Omega_{\rm cdm} + \Omega_{\rm b}$        & $0.2
                   \rightarrow \Omega_{\rm tot} \, (0.1)$ \\
$\Omega_{\Lambda}$ & $ 1 - \Omega_{\rm m}$              &
                   $\Omega_{\rm tot} - \Omega_{\rm m}$ \\
$\Omega_{\rm tot}$     & 1                                & $0.2 \rightarrow 1.0 
\, (0.1)$ \\
$h$                & $0.50 \rightarrow 0.90 \, (0.04)$          & $0.50 
\rightarrow 0.80 \, (0.05)$ 
\\ \tableline
\multicolumn{3}{l}{The step size for independent variables is shown}\\
\multicolumn{3}{l}{in parentheses.}
\end{tabular}
\label{tbl:cosmo}
\end{center}
\end{table*}

We have used the likelihoods of our data to explore limits on
the cosmological parameters shown in Table 1 for both flat and open
model universes with power-law density fluctuation spectra having
slope $n=1$ using CMBFAST \citep{CMBFAST}. $\Omega$ is the density parameter, and subscripts ``${\rm
tot}$,'' ``${\rm b}$,'' ``${\rm m}$,'' ``${\rm cdm}$,'' and ``$\Lambda$`' refer
to the total, baryonic, matter, cold dark matter, and cosmological
constant contributions to the density parameter.  The ranges and
intervals of the parameters are shown in Table 1. For the flat models
we find the likelihood peaks at $\Omega_{\rm b}\, h^2 = 0.009$ and
drops by a factor of 2 at $\Omega_{\rm b}\,h^2=0.019$ 
\citep[e.g.,][]{BURLES,OMEARA}, and by a factor of 3 at $\Omega_{\rm
b}\,h^2=0.03$. For the open models we have assumed uniform priors for
$H_0, \; \Omega_{\rm b} \,h^2, \; \Omega_{\rm cdm}$, and
$\Omega_{\rm tot}$, with $\Omega_{\rm m}$ being uniformly distributed
between 0.2 and $\Omega_{\rm tot}$, over the ranges indicated in Table
1, and we find that $\Omega_{\rm tot} \le 0.4$ or $\Omega_{\rm tot}
\ge 0.7$ at the 90\% confidence level.

\acknowledgements We thank Roger Blandford, Marc Kamionkowski, Andrew Lange, and
Wallace Sargent for useful comments on this work, Russ Keeney for his
tireless efforts on the 40~m telescope, and Angel Otarola for
invaluable assistance in setting up the CBI in Chile.  We gratefully
acknowledge the generous support of Maxine and Ronald Linde, Cecil
and Sally Drinkward, and Barbara and Stanley Rawn, Jr., and the strong
support of the provost and president of the California Institute of
Technology, the PMA division Chairman, the director of the Owens
Valley Radio Observatory, and our colleagues in the PMA
Division. This work was supported by the National Science Foundation
under grants AST 94-13935 and AST 98-02989. We are grateful to CONICYT
for granting permission to operate the CBI at the Chajnantor
Scientific Preserve in Chile.




\end{document}